\documentstyle[epsfig]{mn}
\newif\ifAMStwofonts

\ifoldfss
  \ifCUPmtlplainloaded \else
    \NewTextAlphabet{textbfit} {cmbxti10} {}
    \NewTextAlphabet{textbfss} {cmssbx10} {}
    \NewMathAlphabet{mathbfit} {cmbxti10} {} 
    \NewMathAlphabet{mathbfss} {cmssbx10} {} 
  \fi
  \ifAMStwofonts
    \ifCUPmtlplainloaded \else
      \NewSymbolFont{upmath} {eurm10}
      \NewSymbolFont{AMSa} {msam10}
      \NewMathSymbol{\upi}     {0}{upmath}{19}
      \NewMathSymbol{\umu}     {0}{upmath}{16}
      \NewMathSymbol{\upartial}{0}{upmath}{40}
      \NewMathSymbol{\leqslant}{3}{AMSa}{36}
      \NewMathSymbol{\geqslant}{3}{AMSa}{3E}

      \let\leq=\leqslant 
       \let\ge=\geqslant
    \fi
  \fi
\fi 

\ifnfssone
  \newmathalphabet{\mathit}
  \addtoversion{normal}{\mathit}{cmr}{m}{it}
  \addtoversion{bold}{\mathit}{cmr}{bx}{it}
  \newmathalphabet{\mathbfit} 
  \addtoversion{normal}{\mathbfit}{cmr}{bx}{it}
  \addtoversion{bold}{\mathbfit}{cmr}{bx}{it}
  \newmathalphabet{\mathbfss} 
  \addtoversion{normal}{\mathbfss}{cmss}{bx}{n}
  \addtoversion{bold}{\mathbfss}{cmss}{bx}{n}
  \ifAMStwofonts
    \ifCUPmtlplainloaded \else
      \UseAMStwoboldmath
      \makeatletter
      \new@mathgroup\upmath@group
      \define@mathgroup\mv@normal\upmath@group{eur}{m}{n}
      \define@mathgroup\mv@bold\upmath@group{eur}{b}{n}
      \edef\UPM{\hexnumber\upmath@group}
      \new@mathgroup\amsa@group
      \define@mathgroup\mv@normal\amsa@group{msa}{m}{n}
      \define@mathgroup\mv@bold\amsa@group{msa}{m}{n}
      \edef\AMSa{\hexnumber\amsa@group}
      \makeatother
      \mathchardef\upi="0\UPM19
      \mathchardef\umu="0\UPM16
      \mathchardef\upartial="0\UPM40
      \mathchardef\leqslant="3\AMSa36
      \mathchardef\geqslant="3\AMSa3E

      \let\leq=\leqslant 
       \let\ge=\geqslant
    \fi
  \fi
\fi 

\ifnfsstwo
  \DeclareMathAlphabet{\mathbfit}{OT1}{cmr}{bx}{it}
  \SetMathAlphabet\mathbfit{bold}{OT1}{cmr}{bx}{it}
  \DeclareMathAlphabet{\mathbfss}{OT1}{cmss}{bx}{n}
  \SetMathAlphabet\mathbfss{bold}{OT1}{cmss}{bx}{n}
  \ifAMStwofonts
    \ifCUPmtlplainloaded \else
      \DeclareSymbolFont{UPM}{U}{eur}{m}{n}
      \SetSymbolFont{UPM}{bold}{U}{eur}{b}{n}
      \DeclareSymbolFont{AMSa}{U}{msa}{m}{n}
      \DeclareMathSymbol{\upi}{0}{UPM}{"19}
      \DeclareMathSymbol{\umu}{0}{UPM}{"16}
      \DeclareMathSymbol{\upartial}{0}{UPM}{"40}
      \DeclareMathSymbol{\leqslant}{3}{AMSa}{"36}
      \DeclareMathSymbol{\geqslant}{3}{AMSa}{"3E}

      \let\leq=\leqslant 
       \let\ge=\geqslant
    \fi
  \fi
\fi 

\ifCUPmtlplainloaded \else
  \ifAMStwofonts \else 
    \def\upi{\pi}
    \def\umu{\mu}
    \def\upartial{\partial}
  \fi
\fi

\title{A possible solution to the [$\alpha$/Fe]-$\sigma$ problem in early type galaxies within a hierarchical galaxy formation model}
\author[F. Calura, N. Menci]
       {F. Calura$^{1,2}$\thanks{E-mail: fcalura@oats.inaf.it}, N. Menci$^{3}$\\
        (1) Jeremiah Horrocks Institute for Astrophysics and Supercomputing, University of Central Lancashire, Preston PR1 2HE, UK\\
        (2) INAF, Osservatorio Astronomico di Trieste, Via G.B. Tiepolo
	11, 34131 Trieste, Italy\\
	(3) INAF, Osservatorio Astronomico di Roma, via Frascati 33, I-00040 Monteporzio, Italy\\
	 }
	
\date{Accepted ---- .
      Received ---- ;
      in original form ----}

\pagerange{\pageref{firstpage}--\pageref{lastpage}}
\pubyear{----}

\begin{document}

\maketitle

\label{firstpage}
\begin{abstract}
The most massive elliptical galaxies apparently formed the fastest,
because the ratio of $\alpha$ elements (such as oxygen) to iron is the
smallest. In fact, iron is mainly produced from type Ia supernovae on a timescale of
$\sim 0.1-1$ billion years, while the $\alpha$ elements come from massive stars
on timescales of a few tens of million years (Matteucci 1994). Reproducing such a 
$\alpha$/Fe correlation has long been a severe problem for 
cosmological theories of galaxy formation, which envisage massive galaxies to assemble 
gradually from smaller progenitors, and to be characterized by a star formation history 
too much extended towards late cosmic times. While 
it has recently become clear that feedback from Active Galactic Nuclei (AGN) activity play a
role in the late quenching of star formation (e.g. Cattaneo et al. 2009), and that early star formation history
in the galaxy progenitors affect the $\alpha$/Fe ratio (Calura \& Menci 2009), 
major mergers alone cannot enhance the star formation in the high-redshift progenitors
to the levels required to match the  steepness of the observed $\alpha$/Fe  correlation
(Spolaor et al. 2010). 
Here we report that the inclusion of the effects 
of fly-by 'harassments', that trigger lower 
level starbursts, combined with the AGN quenching of the starburst activity, 
considerably enhances the capability to account for the observed $\alpha$/Fe ratio in ellipticals 
within cosmological galaxy formation models . 
The critical difference between the earlier work and the present result
is the effect of starbursts driven by fly-by encounters that would have
been very common amongst the high-redshift progenitors of massive
galaxies and which would have boosted star formation 
in the first 2 billion years after the Big Bang, 
combined with quenching of the burst activity within the first 3-4 Gyr.  
\end{abstract} 

\begin{keywords}
Galaxies: formation and evolution; Galaxies: abundances.
\end{keywords}

\section{Introduction} 
The correlation between stellar 
$\alpha$/Fe and velocity dispersion observed in local ellipticals (Trager et al. 2000; Thomas et al. 2005; Bernardi et al. 2006; Graves et al. 2007;  
S\'anchez-Bl\'azquez et al. 2006, 
Spolaor et al. 2010; Zhu et al. 2010)  is 
generally referred to as one of the evidences of the 
``downsizing'' pattern of  local galaxies, a pattern which is observable up to high redshift 
(e.g. Cowie et al. 2006; Kodama et al. 2004; Clements et al. 2008) 
 indicating that the more massive the galaxy, the shorter the star formation timescale (Matteucci 1994; Renzini 2006; Pipino et al. 2010; Rogers et al. 2010).  \\
The failure of cosmological galaxy formation models (Thomas et al. 2005; Nagashima et al. 2005; Spolaor et al. 2010) 
in accounting for the above correlation is in fact due to such a  
''downsizing'' aspect. 
In fact, the precision measurements of the power spectrum of the initial density perturbations (Percifal et al. 2007) 
provide evidence for large-scale fluctuations to be smaller, on average, compared to small-scale ones - in accord to what expected 
in a Cold Dark Matter (CDM)  scenario; this implies a hierarchical build-up of dark matter (DM) haloes, where small-mass galaxies collapse first, 
and later assemble to form massive ellipticals (De Lucia et al. 2006). 
The deep physical origin of the disagreement between  the observed and the predicted $\alpha/Fe-\sigma$ correlation 
has been often considered as an evidence for a fundamental flaw of cosmological CDM models of galaxy formation (Thomas et al. 2005; Baugh 2006).

To produce downsizing within standard galaxy formation models, 
some physical mechanisms are required which can enhance star formation 
in massive galaxies at early times, possibly quenching it at late times. 
In cosmological galaxy formation, 
environmental effects are likely to play a major role in producing such a scenario. 
The progenitors of massive galaxies formed from biased regions of the density field, 
and fly-by events and merging in such regions may trigger starbursts at early cosmic times ($t \leq  2.5$ Gyr), 
when the Universe was denser and the rate of galaxy encounters was at least 5-10 times higher than the present. 
A fundamental mechanism to quench star formation in massive galaxies is linked to their 
Active Galactic Nuclei (AGN). 
Feedback from powerful AGN may switch off star formation when the progenitors
assembled into the main progenitor of the final elliptical  (Cattaneo et al. 2009).

These processes may concur in yielding stellar populations in massive galaxies characterized by  
a prompt star formation at early epochs followed by a quenching, in agreement with what 
suggested by the observed  $\alpha/Fe-\sigma$ correlation. Thus, 
testing whether the latter processes may in fact explain the above correlation is  crucial for assessing the consistency of CDM models, 
and requires developed modelling which includes all the physical processed detailed above. \\

In a previous paper (Calura \& Menci 2009, hereinafter CM09), we 
investigated the chemical properties of local galaxies within a cosmological hierarchical clustering scenario  
through a semi-analytic model (SAM) of galaxy formation. 
We used a hierarchical semi-analytic model where chemical evolution was computed 
by taking into account the stellar lifetimes, a significant step forward with respect to the instantaneous recycling approximation.  
However, chemical abundances were computed by considering 
the total (summed over all progenitors) star formation history of each model galaxy
and assumed an average interstellar $\alpha$/Fe. This treatment was suited to the study of interstellar abundances, 
comparable to the ones observed today in stars in the solar neighbourhood, but 
in many cases it lead us to underestimate the integrated stellar 
$\alpha$/Fe abundances measured in local early type galaxies. \\

In this paper, by means of the same SAM, we perform a thorough study of the 
the average stellar $\alpha$/Fe 
in early type galaxies, taking into account the contribution from all the single 
progenitors. Although computationally more expensive, 
this constitutes a step forward with respect to our earlier work (CM09), since thanks to the inclusion 
of the contribution of all the progenitors, our estimates of the integrated abundances are more accurate. 
A second enhancement with respect to our previous work is the calculation of the 
luminosity-weighted abundances (against the mass-weighted ones used in the first paper), another 
aspect which improves the comparison of our results to local observations. \\
The aim of this paper is to show the relative roles of the various processes in shaping the $\alpha/Fe-\sigma$ correlation, 
stressing the importance of some processes which have been neglected in previous studies, such as interaction-triggered
fly-by events and merging-triggered starbursts. \\
This paper is organized as follows. In Sect. 2, our model is briefly described. In Sec. 3, we present our main results, 
which are then discussed in a broader context in Sect. 4. 

\section{The model}
To study the integrated abundances of local ellipticals, 
we use the same model as described in Calura \& Menci (2009). 
We defer the reader to that paper for further details. Here we summarize the main features of our model. \\
Galaxy formation is driven by the collapse and growth of DM halos, which originate by gravitational instability of  overdense regions in the 
primordial DM density field, a random,
Gaussian  density field with CDM power spectrum within the
''concordance cosmology" (Spergel et al. 2007) for which we adopt round
parameters  $\Omega_{\Lambda}=0.7$, $\Omega_{0}=0.3$, baryonic density
$\Omega_b=0.04$ and Hubble constant (in units of 100 km/s/Mpc) $h=0.7$.
As  cosmic time increases, larger and larger regions of the density field
collapse, and  eventually lead to the formation of groups and 
clusters of galaxies, which grow by merging with mass and redshift dependent rates
provided by the Extended Press \& Schechter formalism (Press \& Schechter 1974). \\
The clumps included into larger DM halos 
may survive as satellites, merge to form larger galaxies due to binary 
aggregations,  or coalesce into the central dominant galaxy due to dynamical 
friction. The above complex dynamical evolution of galaxies is followed through a 
Monte Carlo simulation  of the collapse and subsequent merging history of the peaks of the 
primordial density field, which enables to generate a synthetic catalogue of model galaxies and of 
their past merging history. \\
For each model galaxy, we follow the evolution of its gas and stellar content; 
the gas which has radiatively cooled in the galactic haloes (with mass $M_g$) 
partly condenses into stars at a rate $\psi \propto M_g/t_d$ on long timescales,  
$t_d\approx 2-10$ Gyr according to the Kennicutt law (Kennicutt 1998) . 
Star formation can also occur  through 
interaction-triggered starbursts (ITS), driven by merging or 
by fly-by events between galaxies.
Such a star formation mode can convert large fractions (up to 100 \% in major merging events) 
of the cold gas on short timescales ($\approx 10^6$ yrs), and 
provides an important contribution to the early formation of stars
in massive galaxies (Menci et al. 2004a). \\
The model also includes a detailed treatment of the growth of supermassive
black holes at the centre of galaxies, of the corresponding AGN activity powered by the 
 interaction-triggered inflow
of cold gas onto the black holes, and of the AGN feedback onto the galactic gas. 
Due to the crucial role of AGN feedback in the affecting the star formation history of 
massive galaxies, this sector of the model has been extensive tested; the predicted QSO luminosity function and 
X-ray luminosity function and emissivity have been checked against observation over a wide range of redshift 
$0\leq z\leq 5$ and bolometric luminosities $10^{43}\leq L/erg s^{-1}\leq 10^{46}$ (Menci et al. 2003; 2004a); the description of the AGN feedback 
as resulting from the sweeping of galactic gas by the blast wave produced by the injection of the AGN radiative energy has 
been tested by comparing the absorption properties of AGNs (as a function of both luminosity and redshift) with recent observation
(Menci et al. 2008); the consistency of the modelling of the AGN feedback with the color distribution of 
galaxies has been discussed in Menci et al. (2006).\\
As shown in Fontana et al. (2004), the inclusion of processes such as ITS and AGN feedback within the SAM 
affects mostly the  model rendition of the stellar mass function at high redshift. 
All models discussed in this paper have basically the same stellar mass function at $z \sim 0$. \\
Galaxy luminosities in different bands are computed by convolving the 
computed star formation history (SFH) with synthetic spectral libraries of stellar populations (Bruzual \& Charlot 2003). \\
To compare our prediction to the observed $\alpha$/Fe-$\sigma$ relation, 
as performed in our previous paper, we select elliptical galaxies 
from the SAM synthetic catalogue on the basis of their present-day (B-V) colour. 
In specific, we consider elliptical galaxies all the systems with present (B-V)$\ge 0.85$ (Roberts \& Haynes 1994). 
For any selected galaxy, we compute the chemical evolution a posteriori by means 
of detailed chemical evolution equations taking into account the stellar lifetimes, 
fundamental for studying the chemical evolution of Fe. 
Thus, the SFH, the gas content, and the gas inflow and outflow histories of all the progenitors of the considered 
galaxy provided by the SAM constitute the input for calculating the chemical evolution. 
This constitutes a step forward with respect to our earlier work (CM09), 
where we considered only the total (summed over all progenitors) star formation history of each model galaxy
and assumed an average interstellar $\alpha$/Fe. 

\section{Results}
\begin{figure*}
\centering
\vspace{0.001cm}
\epsfig{file=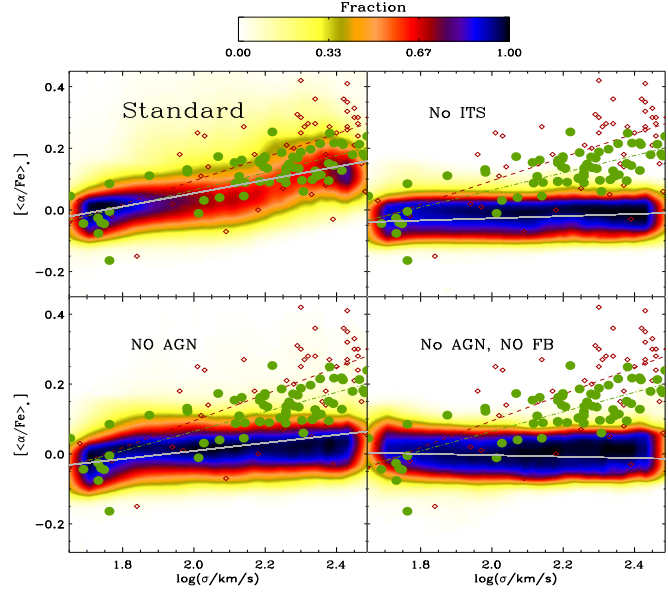,height=8cm,width=9cm}
\caption{Stellar average $\alpha/Fe$ ratio vs velocity dispersion compared to local observations. 
The colour code represents the predicted number of galaxies with a given $\alpha/Fe$ 
and a velocity dispersion  $\sigma$, 
normalised to the total number of galaxies with  that  velocity dispersion. 
The solid lines represent  linear-regression fits to the model. 
The dashed line and dash-dotted  line  represent  linear-regression fits to the observational   
$\alpha/Fe$-$\sigma$ relations, as compiled by Spolaor et al. (2010; open diamonds) and Arrigoni et al. (2009; solid circles)
respectively.  
\emph{Top-left panel}: our results have been computed by means of our standard assumptions, i.e. by taking into account both interaction-triggered 
starbusts at high redshift and AGN feedback. \emph{Top-right panel}: no interaction-triggered starbursts and fly-by interactions.   
\emph{Bottom-left panel}: no AGN feedback.  \emph{Bottom-right panel}: no AGN feedback, no  fly-by interactions.  }
\label{fig1}
\end{figure*}

The average stellar $\alpha/Fe$ may be calculated as: 
\begin{equation}
<\alpha/Fe>_{*}\simeq \frac{\int (\alpha/Fe)(t) \psi(t) L^{SSP}_V(t_0-t)dt}{\int \psi(t) L_V(t_0-t)dt}
\label{afe}
\end{equation}
i.e. it is the time integral of the interstellar $(\alpha/Fe)$ abundance over 
the star formation history, represented by the star formation rate $\psi(t)$, multiplied by the 
luminosity $L^{SSP}_V(t_0-t)$, which is the luminosity at epoch $t_0$ of a simple stellar population
 of unitary mass with an age of $(t_0-t)$, and is given by SSP photometric models as a function of age and metallicity (BC03),  
divided by the present total luminosity.\\
In Fig.~\ref{fig1}, we show the predicted $\alpha/Fe$ vs $\sigma$ relation for our selected elliptical galaxies, considering four 
different cases.  
In Fig.~\ref{fig1}, we report two different sets of data (Arrigoni et al. 2009; Spolaor et al. 2010), 
in order to have an idea on the uncertainty of 
 the slope of the observed $\alpha$/Fe-$\sigma$ relation. 
In the standard case, we consider a model including interaction-triggered starbursts in the most massive galaxies and Active Galactic Nuclei 
(AGN) feedback.  
In the second case, we suppressed the ITS and consider AGN feedback alone. 
In the third case, we suppressed AGN feedback and consider  ITS alone. 
Since the ITS consist of two components (merging and fly-by events), it may be interesting to separate these two processes in order to 
appreciate the importance of fly-by events. 
In the fourth case, we show our results obtained having suppressed AGN feedback and fly-by interactions, but considering only merger-triggered starbursts. 
In the standard case, the slope of the observed $\alpha/Fe$-$\sigma$ relation  is satisfactorily reproduced. 
In the second and third cases, although with our model a correlation between $\alpha/Fe$ and $\sigma$ is obtained, 
the predicted relation is considerably shallower than the observed one. 
In the fourth case, the predicted $\alpha/Fe$-$\sigma$ relation is overly flat with respect to the observations.  
The comparison between the third and fourth cases is  essential to understand the major role of fly-by interactions 
in driving the 
correlation between $\alpha$/Fe and $\sigma$.

\section{Discussion}

Our results show that a standard CDM  galaxy formation model including  ITS, 
consisting in both merger-triggered and fly-by triggered starbursts, 
and AGN feedback can naturally lead 
to shorter star formation timescales in larger galaxies, i.e. to
star formation histories reversed with respect to their mass-assembly histories.\\
The absolute novel ingredient of our model consists in the simultaneous inclusion of both 
the starbursts triggered by fy-by events, absent in any previous investigation of this 
fundamental property of early-type galaxies, and the AGN feedback related to the active Quasar phase
of AGNs.\\
The general problem of quenching the star formation histories in massive galaxies have been investigated in several 
studies (e.g. Gabor et al. 2010) and is of great interest. 
Fig. 2 is useful to better understand how the features of our model act directly on the star formation histories 
of single galaxies. 
In each panel of this figure, we show  the SFH of three galaxies 
drawn from the sample including ITS and AGN, including AGN but not ITS and from the one including ITS but not the effects of AGNs. 
The three galaxies shown in each panel present the same merger histories. It is worth to stress that in each case, the final 
stellar masses are not the same, since the inclusion of physical ingredient such as ITS and AGN can considerably alter the 
star formation hisory and  the feedback history of the systems. The final stellar mass of each model is reported in  
Fig. 2 (see caption for further details).\\
It is important to note that the suppression of ITS has effects also on the efficiency of the AGN feedback, 
since, according to the model used here and described in Menci et al. (2008), the cold gas which causes the 
interact-triggered starbursts feeds also the nuclear activity. In fact, in every panel, 
both  models without ITS and AGN 
present higher initial star fromation rates with respect to the standard case. 
The explanation of this is that  in both cases (no ITS and no AGN) the 
AGN is inefficient in heating and ejecting gas from the galaxies, resulting in higher amount of cold gas when the star formation 
has already turned on.\\
At the present time, the models which include ITS but without AGN feedback  show SFR values larger than those shown by the 
ones drawn from the standard sample. 
Globally, their SFHs are less skewed towards early times with respect to the ones of the model with both AGN and ITS. 
The effects of AGNs in quenching the star formation activity within  the first 3-4 Gyr is evident from this figure. 
Finally, the model with AGN but without ITS presents very extended SFHs, with rather flat and smooth behaviours and 
very high present-day values. 
Without the effect of ITS, gas consumption at early times is not very efficient and at later times, after a few Gyrs, 
the large amounts of cold star-forming gas prevent AGNs from being effective and quenching the star formation. 
It is therefore clear that, in order to have realistic star formation histories for early type galaxies, 
both effects of ITS and AGN feedback are necessary.\\
Both effects of AGN feedback and ITS seem to play an important role in determining galactic downsizing. 
However, the effect of the former seems slightly dominant, 
and this is supported by the fact that the predicted $\alpha/Fe$-$\sigma$ relation without ITS is slightly shallower than 
the one without AGN. \\
Our study of  the $\alpha/Fe$-$\sigma$ relation tells us that our galaxy formation model can naturally 
account for one of the most important aspects of the  ``downsizing'' character of early type galaxies. 
Downsizing (i.e. oldest stars are in most massive galaxies, star formation shifts to lower 
masses at late times) in hierarchical models will arise if there is a minimum mass 
required to have star formation and a maximum mass above which feedback from AGN  
suppresses further star formation (Sheth 2003; Croton \& Farrar 2008).  
Together, these will 
also lead to an [$\alpha$/Fe]-sigma relation, since in principle, in such a scenario, 
the distribution of the formation timescales will be narrower for the stars forming in the most 
massive systems. 
In this framework, ITS help because 
it allows to use up more of the gas earlier, thus leaving less work to be done by the AGN feedback, 
and leading to a steeper [alpha/Fe]-sigma relation.  A detailed calculation of chemical abundances 
in more hirarchical  models  (such as the ones above) will be of further help in understanding the relevance of these processes 
with respect to the downsizing character of local galaxies.  \\
It is important to stress that 
the observations can be reproduced without 
the modification of any fundamental 
chemical evolution parameter,  such as the stellar IMF. 
However, it is worth to stress that such ingredient may play some role in determing the shape of 
the $\alpha/Fe$-$\sigma$ relation. 
In fact, 
Although our unprecedented rendition of this relation, the predicted slope obtained with our standard model is slightly shallower 
than the observed one. 
This aspect will be investigated in our future 
work, and may have its explanation in effects such as a possible IMF dependence on the SFR and/or a flatter IMF in starbursts 
(Recchi et al. 2009; Haas \& Anders 2010; Calura et al. 2010). 
Some direct evidences in favour of the latter hypotesis come from observational studies of nuclear star clusters in the Milky Way 
(Bartko et al. 2010). Also dynamical investigations of early-type galaxies 
seem to require stronger baryonic feedback at the epoch of the formation of these systems, 
fully consistent with an IMF skewed towards more massive stars (Napolitano et al. 2010), 
or from the study of local ultra-compact galaxies (Dabringhausen et al. 2009). A top-heavy IMF at early times seems also 
required in order to reproduce the abundance ratios observed in the hot intracluster medium 
(Matteucci \& Gibson 1995; Loewenstein \& Mushotzky 1996).\\
In our previous work (C09), 
the average stellar abundance ratios have been calculated 
without 
taking into account the contribution of the single 
progenitors, i.e. by considering the total star formation history of the selected galaxy, given by the sum of the star formation rates 
of all the progenitors, and considering an average interstellar $\alpha$/Fe. 
In that paper,  the observed $\alpha$-Fe-$\sigma$ relation could be partially explained 
by 
assuming a top-heavy IMF in the most massive galaxies,  
which mimics the effect of a higher star-formation efficiency (or, in other words, 
a shorter star formation timescale) in largest galaxies (Matteucci 1994; Ferreras \& Silk 2003; Matteucci et al. 1998).\\
The largest galaxies have the most massive progenitors, which form all their stars in the shortest timescales 
and dominate the integral of Eq.~\ref{afe}. 
This is the main reason why here, by considering the SFRs and the abundance ratios 
of each single progenitor, instead of a total SFR and an abundance  averaged over all the progenitors, 
we can successfully reproduce the observed $\alpha/Fe$-$\sigma$ relation without any modification of the main chemical 
evolution parameters. \\


\begin{figure*}
\centering
\vspace{0.001cm}
\epsfig{file=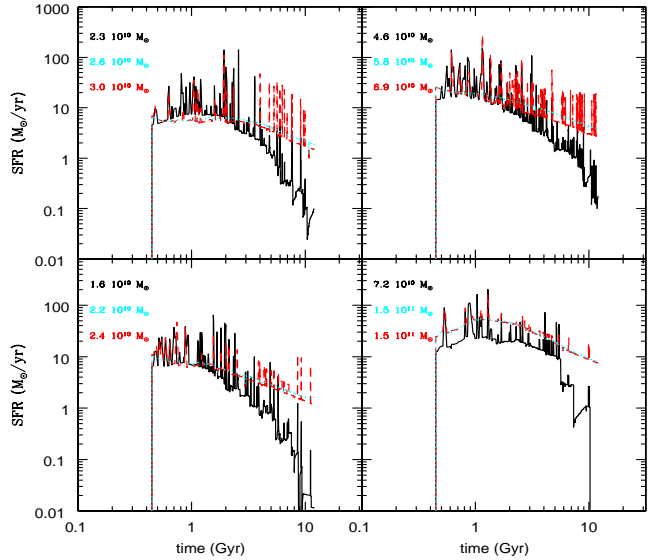,height=8cm,width=9cm}
\caption{Star Formation histories of a few early-type galaxies of our model. In each panel, we show 
the SFH of three galaxies  drawn from the sample including ITS and AGN (solid black lines), 
including AGN but not ITS (cyan dotted  lines) 
and including ITS but not AGNs (dashed red slines). In each panel, the three systems present the same merging history. 
The final stellar masses of the models  are indicated in each panel. 
From top to bottom, the indicated masses refer to the standard case, the No ITS case and the No AGN case. 
}
\label{fig2}
\end{figure*}

\section*{Acknowledgments}
FC wish to acknowledge several interesting discussions with F. Matteucci, B. K. Gibson and P. Sanchez-Blazquez. 
We also acknowledge useful comments by an anonymous referee.

\label{lastpage}

\end{document}